\documentclass[twocolumn]{aastex631}
\usepackage{relsize} % for \mathlarger
\usepackage{graphicx}	% Including figure files
\usepackage{amsmath}	% Advanced maths commands
\usepackage[flushleft]{threeparttable}
\usepackage{changepage}
\usepackage{epstopdf}
\usepackage{amsopn,amsxtra,txfonts}
\usepackage{comment} 
\usepackage{longtable,threeparttablex}
\usepackage{enumitem}
\usepackage{fancyhdr}
\usepackage{booktabs}
\usepackage[hang,flushmargin]{footmisc}
\usepackage{gensymb}
\usepackage{tabularx}
\usepackage{ragged2e}
\usepackage{float}
\usepackage{enumitem}
\usepackage{hyperref}
\usepackage{wrapfig}
\usepackage{sidecap}
\usepackage{tcolorbox} % Ensure this package is installed
\usepackage{xcolor} % For color definitions
\usepackage{natbib}
\usepackage{threeparttable}
\usepackage{array}
\usepackage{booktabs}

\newcommand{\NpredAGN}{log\,$N^{\rm pred}_{\text{\tiny AGN}}$}

\newcommand{\dellogN}{$\Delta$log\,$N$}
\newcommand{\caldef}{\ion{Ca}{2} deficit}
\newcommand{\Wsens}{$W^{\rm sens}_{3\sigma}$}
\newcommand{\Nsens}{log\,$N^{\rm sens}_{3\sigma}$}

\newcommand{\Wmeas}{$W^{\rm meas}$}
\newcommand{\Nmeas}{log\,$N^{\rm meas}_{3\sigma}$}
\newcommand{\tn}{\tablenotetext}
\newcommand{\CaII}{\ion{Ca}{2}}
\newcommand{\NaI}{\ion{Na}{1}}
\newcommand{\HI}{\ion{H}{1}}
\newcommand{\hcop}{\text{HCO\textsuperscript{+}}}

\newcommand{\logNHI}{log\,(N(\ion{H}{1})/cm$^{-2}$)}

\newcommand{\msun}{\,$\rm M_{\odot}$}

\newcommand{\kms}{\,km\,s$^{-1}$}

\begin{document}

%\title{The Magellanic Stream Distance: Insights from Optical Absorption lines along Stellar sightlines}
\title{The Distance to the Magellanic Stream: Constraints from Optical Absorption along Stellar Sightlines}

\author[0000-0002-4157-5164]{Sapna Mishra}
\affiliation{Space Telescope Science Institute, 3700 San Martin Drive, Baltimore, MD 21218, USA}

\author[0000-0003-0724-4115]{Andrew J. Fox}
\affiliation{AURA for ESA, Space Telescope Science Institute, 3700 San Martin Drive, Baltimore, MD 21218}
\affiliation{Department of Physics \& Astronomy, Johns Hopkins University, 3400 N. Charles Street, Baltimore, MD 21218, USA}

\author[0000-0002-3082-7266]{J. V. Smoker}
\affiliation{European Southern Observatory, Alonso de Cordova 3107, Vitacura, Santiago, Chile}
%\affiliation{UK Astronomy Technology Centre, Royal Observatory, Blackford Hill,Edinburgh EH9 3HJ,UK}

\author[0000-0001-9982-0241]{Scott Lucchini}
\affiliation{Center for Astrophysics | Harvard \& Smithsonian, 60 Garden Street, Cambridge, MA 02138, USA}
\affiliation{Department of Physics, University of Wisconsin- Madison, Madison, WI 53706, USA}

\author[0000-0003-2676-8344]{Elena D'Onghia}
\affiliation{Department of Physics, University of Wisconsin- Madison, Madison, WI 53706, USA}
\affiliation{Department of Astronomy, University of Wisconsin- Madison, Madison, WI 53706, USA}
\affiliation{INAF - Osservatorio Astrofisico di Torino, via Osservatorio 20, 10025 Pino Torinese (TO), Italy}

\correspondingauthor{Sapna Mishra}
\email{smishra@stsci.edu}

\begin{abstract}

The Magellanic Stream (MS) is a large tail of neutral and ionized gas originating from tidal and hydrodynamical interactions between the Magellanic 
Clouds as they orbit the Milky Way (MW). It carries a significant gas reservoir that could impact the future evolution of the MW. Despite its importance, no direct observational constraints on the Stream's distance have been previously published. In this study, we analyze Very Large Telescope/Ultraviolet and Visual Echelle Spectrograph (VLT/UVES) spectra of five blue horizontal branch (BHB) stars in the MW halo located at distances ranging from 13 to 56 kpc near two regions of the Stream (near Stream longitudes of $-$79$^\circ$ and $-$98$^\circ$), with the aim of detecting \CaII\ and \NaI\ absorption. No \CaII\ or \NaI\ absorption is detected at Stream velocities in any of the individual spectra, or in higher signal-to-noise stacks of the spectra. The resulting limits on the \CaII\ absorption are significantly lower than the \CaII\ columns measured in the Stream along extragalactic sightlines. These non-detections establish a firm lower distance limit of 42 kpc for one region of the Stream. For the other region, we set a firm lower limit of 20 kpc and a tentative lower limit of 55 kpc from the most distant star, but deeper spectra are needed to confirm this. Our results provide the first observational constraints on the gaseous Stream's distance. 

\end{abstract}


\section{Introduction}
\label{sec:intro}

The Magellanic Stream (MS, also referred as the Stream) is an extensive and intricate filamentary structure of neutral and ionized gas within the circumgalactic medium (CGM) of the Milky Way (MW), stretching over 200 degrees across the sky \citep{Nidever2010}. It is thought to originate from tidal and hydrodynamical interactions between the Large and Small Magellanic Clouds (LMC and SMC) as they orbit the MW \citep[see][and references therein]{Elena2016}. Due to its proximity to us, the Stream serves as a rich reservoir of information about galaxy interactions and has been a focal point of study across various disciplines. Observationally, researchers have investigated the morphology and kinematics of the Stream in \HI\ emission \citep{Mathewson1974,Putman2003,Nidever2008,McClure-Griffiths2009}, explored its chemical and ionization properties \citep{Fox2010,Fox2013,Richter2013,Fox2014}, utilized the Stream to constrain the orbital history of the Magellanic Clouds \citep{Kallivayalil2006}, and studied its stellar components \citep{Zaritsky2020, Zaritsky2024, Chandra2023}. Additionally, the Stream has been extensively investigated through simulations to constrain the interaction and passage history of the Magellanic Clouds around the MW \citep{Besla2012, Pardy2018, Lucchini2021, Lucchini2024,Carr2024}.\par

The Stream has important implications for the future evolution of the MW. \citet{Fox2014} estimated that the trailing Stream carries over 10$^{9}$\msun\ of gas, making it a significant source of fuel for future star formation in the MW. This is also evident from the gas accreting onto the MW disk from the leading tail of the Stream, known as the Leading Arm \citep{Fox2018,Tepper-Garcia2019}. However, it remains uncertain when and whether this gas will settle onto the Galactic disk, a process that also depends on the proximity of the Stream to the MW disk.

The distance to the gas in the Leading Arm of the Stream has been observationally constrained using \HI\ kinematics to be $\sim$20 kpc \citep{McClure-Griffiths2008, AntwiDanso2020}. Additionally, recent studies by \citet{Price-Whelan2019} and \citet{Nidever2019}, based on the discovery of Price-Whelan 1 -- a young, low-mass, metal-poor stellar association -- have placed the LA II at a heliocentric distance of $\approx$28 kpc. For the trailing Stream, the distance to the \emph{stellar} component is estimated to range between 60 and 120 kpc, \citep{Chandra2023, Zaritsky2024}, but the distance to the gaseous Stream is observationally unconstrained.\par
Simulations suggest that the gaseous Stream lies between 20 and 150 kpc \citep[][see also review by \citet{Lucchini2024}]{Besla2012, Pardy2018, Lucchini2021}. In the original first passage simulation from \citet{Besla2012}, the Stream is formed outside the MW through dwarf-dwarf interactions. This placed the Stream out at distances exceeding 100 kpc from the Sun. Subsequent models introduced alternatives, accounting for the presence of the Magellanic Corona. Recent estimates suggest the LMC has a mass of $\gtrapprox$10$^{11}$\msun\ \citep[][]{Watkins2024}, leading to the concept of the ``Magellanic Corona"—a warm-hot (T $\approx$ 10$^{5}$ K) halo surrounding the LMC that originally extended to its virial radius \citep{Lucchini2020, Lucchini2021}. This Magellanic Corona has been detected in UV absorption-line observations of high ions \citep{Krishnarao2022} and low ions \citep{Mishra2024b}. Incorporating the Corona into hydrodynamical simulations of the Magellanic Clouds helps resolve several observational puzzles, including the high mass of ionized gas \citep{Fox2014} in the Stream and the survival of the Leading Arm \citep{Lucchini2020}. The more recent model by \citet{Lucchini2021}, with an updated first-passage interaction of the Magellanic Clouds, and including both Galactic and Magellanic Coronae, places the Stream much closer—at just 20--30 kpc from the Sun. However, direct observational constraints on the Stream's distance are needed to test this.\par

UV and optical absorption-line spectroscopy using extragalactic background sources such as active galactic nuclei (AGNs) has been widely employed  to constrain the chemical and kinematic properties of the gas in the Stream \citep{Songaila1981, Lu1998, Gibson2000, Sembach2001, Fox2010, Fox2013, Fox2014, Richter2013}. The observed chemical and kinematic bifurcation of the Stream into two filaments suggest that it originates from both Magellanic Clouds. While these studies impose strong constraints on the properties of Stream gas, its distance remains undetermined observationally. The most promising method for estimating the distance to the Stream's gas is absorption-line spectroscopy of stars in the MW halo. \citet{Lehner2012,Lehner2022} found an absence of very high velocity clouds (VHVCs, $|v_{LSR}|\gtrsim170$\kms) in the UV spectra of MW halo stars at distances of 5--15 kpc. Since the Stream is an example of a VHVC, this indicates that the Stream lies beyond 5--15 kpc of the Sun.
%are associated with the Magellanic Stream and lie within 5–15 kpc of the Sun.} 
Recently, \citet{Steffes2024} used 10 stellar sightlines passing through the Stream to search for cold molecular gas in the Stream. The authors found no detectable cold gas traced by \hcop, HCN, HNC, and C$_2$H in the Atacama Large Millimeter Array (ALMA) data of these 10 stars, leading them to explore various possible explanations for the non-detection. However, their study did not place any constraints on the Stream's distance.\par

Motivated by the need to better constrain the distance to the gas in the trailing Stream, this paper presents a spectroscopic analysis of five blue horizontal branch (BHB) stars in the MW halo that pass through the trailing tip of the Stream from two regions. Using high-resolution optical spectra from the Very Large Telescope/Ultraviolet and Visual Echelle Spectrograph \citep[VLT/UVES,][]{Dekker2000}, we examine the \CaII\ and \NaI\ absorption features in these stars, which provide crucial insights into the Stream's distance. This paper is structured as follows: In Section~\ref{sec:sample_ana}, we describe our sample selection, observations, and data preparation. Section~\ref{sec:method} details our analysis, including the method used to estimate distances using stars, and our measurement and stacking analysis techniques. We present our results in Section~\ref{sec:results}, followed by a discussion and summary of our findings in Section~\ref{section:discussion} and Section~\ref{sec:summary} respectively.\par

\section{Sample and Data Handling} \label{sec:sample_ana}

\subsection{Sample Selection}
\label{subsec:sample}

%%%--------------- Table 1  (AGNs) -----------------
\begin{deluxetable}{lllllll}
\label{tab:agn_summary}
\tabletypesize{\footnotesize}
\tablewidth{0pt}\tabcolsep=2.2pt
\tablecaption{Published MS \CaII\ and \NaI\ Measurements along AGN Sightlines}
\tablehead{AGN  & log\,$N$\,(\HI) & \multicolumn{2}{c}{log\,N\,(\CaII)}      &  \multicolumn{2}{c}{log\,$N$\,(\NaI)}   &  Ref.  \\
                &              &    3934~$\AA$         &    3969~$\AA$    &   5891~$\AA$          &  5897~$\AA$    &        \\       
                & (cm$^{-2}$)  &    (cm$^{-2}$)        &    (cm$^{-2}$)   &  (cm$^{-2}$)          &  (cm$^{-2}$)   &         \\
(1)             &   (2)        &   (3)                 &   (4)            &  (5)                  & (6)            &  (7)    }
\startdata
FAIRALL\,9       &  19.95      &  12.38$\pm$0.03$^{\dagger}$ &  -                  &  11.33$\pm$0.05$^{\dagger}$  & -             &   (a)  \\     
RBS\,144           &  20.17      &  12.27$\pm$0.02             &  12.34$\pm$0.03     &  $<$11.12                    & $<$11.42      &   (b)   \\     
HE\,0056$-$3622  &  18.70      &  11.99$\pm$0.04             &  12.13$\pm$0.06     &  $<$11.37                    & $<$11.63      &   (b)   \\   
PHL\,2525        &  $<$18.21   &  $<$11.19$^{\star}$         &  $<$11.34$^{\star}$ &  $<$10.38$^{\star}$          & -             &   (c)   \\   
NGC\,7469        &  18.63      &  11.43$\pm$0.08             &  11.49$\pm$0.08     &  $<$11.12                    & -             &   (d)  \\
\enddata
\tablecomments{(1) AGN name; (2) \HI\ column density from the GASS survey \citep{McClure-Griffiths2009}; (3) and (4) column densities of \CaII\ 3934,3969 doublet lines; (5) and (6) column densities of and \NaI\ 5891,5897 doublet lines; (7) References: (a) \citet{Richter2013}; (b) \citet{Fox2013}; (c) This work; (d) \citet{Fox2010}. All \CaII\ and \NaI\ column density measurements were made using the Apparent Optical Depth (AOD) method.}\tn{\dagger}{Column density from fit to both doublet lines.} \tn{\star}{ We estimated the AOD upper limits for these lines at the Stream velocity of $v_{\rm LSR}=-255$\kms\ using the spectral data from \citet{Fox2013}.}
\end{deluxetable}
%%%%---------------

%%%%------------- Fig. 1 sample-----------
\begin{figure}[!hbt]
\hspace{-0.5in}
\includegraphics[width=0.6\textwidth]{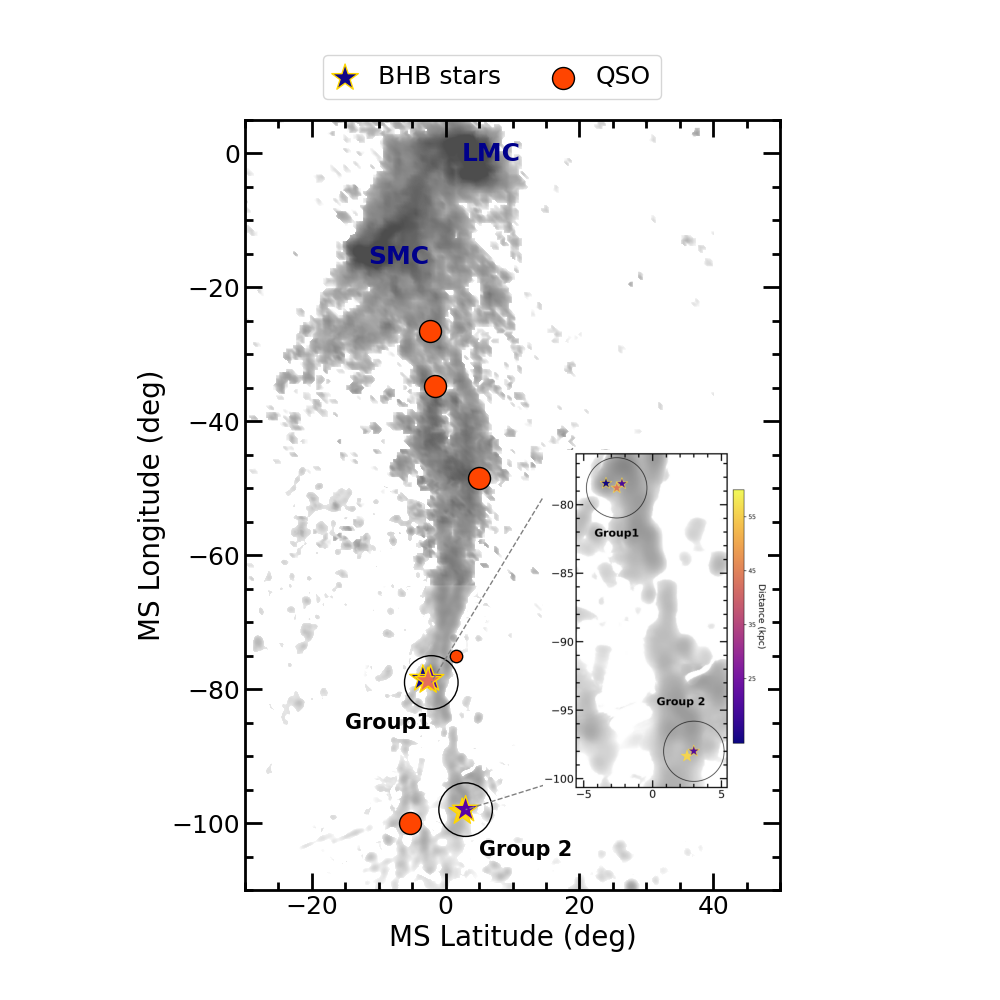}
 \caption{\HI\ 21 cm map of the Magellanic System from GASS survey \citep{McClure-Griffiths2009}. The map is presented in Magellanic Stream coordinates and color-coded by the \HI\ column density. The locations of the Large Magellanic Cloud (LMC) and Small Magellanic Cloud (SMC) are marked for visual reference. The positions of the BHB stars included in this study, situated in two regions of the Stream with \logNHI~$>$19, and categorized as Group 1 and Group 2, are shown in stars. An inset plot provides a zoomed-in view, color-coded by their distances, of these stars along the two groups. For comparison, the locations of AGN sightlines probing the Stream (with \logNHI~ $>$ 18.5), as studied by \citet{Fox2013} and \citet{Richter2013}, are highlighted with red circles. An AGN sightline probing the Stream with \logNHI~ $<$ 18.2 from \citet{Fox2010}, is shown with a smaller red circle.}
 \label{fig:sample} 
\end{figure}
%%%--------------

To determine the distance to the gas in the Magellanic Stream, we look for interstellar absorption in the spectra of background stars with known distances. The most suitable stars for this purpose are blue horizontal branch (BHB) and RR Lyrae stars. In this study, we select BHB stars as background targets because they are non-variable, bright and blue. Furthermore they have well-defined continua that are straightforward to model with minimal stellar lines, and have accurate spectroscopic distance and velocity estimates. To identify a sample of MW halo stars, we used the catalog of BHB stars compiled by \citet{Xue2011}. This catalog contains approximately 4,000 BHB stars from the Sloan Digital Sky Survey Data Release 8 \citep[SDSS DR8,][]{Aihara2011}. The distances of these stars from the Sun in this catalog are accurate to within 5\%.
\par

Next, we concentrate on high-column density regions of the Stream with \HI\ column densities \logNHI $\ge$ 19, as derived from 21 cm \HI\ emission observations from the Parkes Galactic All-Sky Survey \citep[GASS,][]{McClure-Griffiths2009}. This survey has an effective angular resolution of approximately 16$^{'}$ and a velocity resolution of 1.0 \kms, and hence represents the most sensitive and highest angular resolution survey of Galactic \HI\ emission of the Southern sky. This \HI\ column density cut is informed by extragalactic studies that examine Stream properties using \CaII\ and \NaI\ absorption lines in the optical spectra of background AGNs  \citep[][]{Fox2010,Fox2013,Richter2013}. This cut ensures that the Stream regions possess sufficient \HI\ column density to detect \CaII\ and \NaI\ absorption from the Stream (see Table~\ref{tab:agn_summary} for the Stream measurements from AGNs). By cross-correlating the Stream regions of \logNHI\ $\ge$ 19 with the BHB catalog, we identified a sample of 5 BHB stars at distances ranging from 13 to 56 kpc, projected onto two specific regions of the Stream designated as
Group 1 (l$_{\rm MS}$, b$_{\rm MS}$ $\approx$ $-$79\degree, $-$2\degree) and Group 2 (l$_{\rm MS}$, b$_{\rm MS}$ $\approx$ $-$98\degree, 3\degree). The positions of these stars in these two groups on the \HI\ emission map are shown in Fig.~\ref{fig:sample}. The two groups are separated by approximately 20 degrees, with each group including at least one nearby star ($\approx$20 kpc) and one distant star ($>$40 kpc) projected close together on the sky. Including two separate regions of the Stream is an important part of our experimental design, as we can look for consistent distance measurements from both regions. The locations of existing AGN sightlines that trace the Stream region with comparable column densities are indicated by red circles in Fig.~\ref{fig:sample}. The properties of the stars and the surrounding Stream are summarized in Table~\ref{tab:sample_properties}.\par

%%%%============== Table 2 =========
\begin{deluxetable*}{llllllllll}
\label{tab:sample_properties}
\tabletypesize{\footnotesize}
\tablewidth{0pt}\tabcolsep=4pt
\tablecaption{Basic Properties of the Stellar Sightlines}

\tablehead{Star Name                 &   GLON  &   GLAT      &  $g$-mag  &  $d$      &  $v_{\rm rad}$         &   log\,$N$\,(\HI)      &  $v_{\rm LSR}$(\HI) &    Group   &   Texp   \\
                          &   (deg) &   (deg)     &         &  (kpc)  &  (km s$^{-1}$)           &  (cm$^{-2}$)   & (km s$^{-1}$)    &            &   (hr)   \\
(1)			              &   (2)   &   (3)       &  (4)    &   (5)   &  (6)                     &  (7)           &   (8)            &    (9)     &   (10)   \\}
\startdata
SDSSJ233702.19$-$110249.3  &   71.94 &  $-$66.24   &  16.22  &  13.3$\pm$0.7  &     236     $\pm$ 1.6    &  19.57  &   $-$220  &    Group1  &   2      \\    
SDSSJ234130.32$-$104009.1  &   74.59 &  $-$66.73   &  16.97  &  19.3$\pm$1.0  &   $-$156  4 $\pm$ 2.0    &  19.36  &   $-$218  &    Group1  &   2      \\    
SDSSJ233940.64$-$102939.6  &   74.06 &  $-$66.29   &  18.76  &  42.8$\pm$2.1  &   $-$316  6 $\pm$ 4.2    &  19.14  &   $-$212  &    Group1  &   8      \\    
SDSSJ233736.67+093011.1	   &   94.37 &  $-$49.23   &  17.16  &  20.5$\pm$1.0  &   $-$168  6 $\pm$ 6.2    &  19.08  &   $-$323  &    Group2  &   2      \\    
SDSSJ233516.61+094216.4	   &   93.71 &  $-$48.80   &  19.29  &  55.9$\pm$2.8  &   $-$139  0 $\pm$ 13.9   &  18.98  &   $-$325  &    Group2  &   8      \\    
\enddata
\tablecomments{Col. (1) BHB star name; Cols (2) and (3) Galactic longitude and latitude; Col. (4) Extinction-corrected $g$-band magnitude; Col. (5) Distance from the Sun; Col. (6) Heliocentric stellar radial velocity; Cols (7)  and (8) \HI\ emission column density and central LSR velocity of the MS in the direction of the BHB stars from GASS survey; Col. (9) Group classification; Col. (10) Exposure time for VLT/UVES observations. The values in columns 1–6 are taken from \citet{Xue2011}.}
\end{deluxetable*}
%%%%---------------
%%%%------------- Fig. 2 Flow chart -----------
\begin{figure*}
 \includegraphics[width=0.9\textwidth]{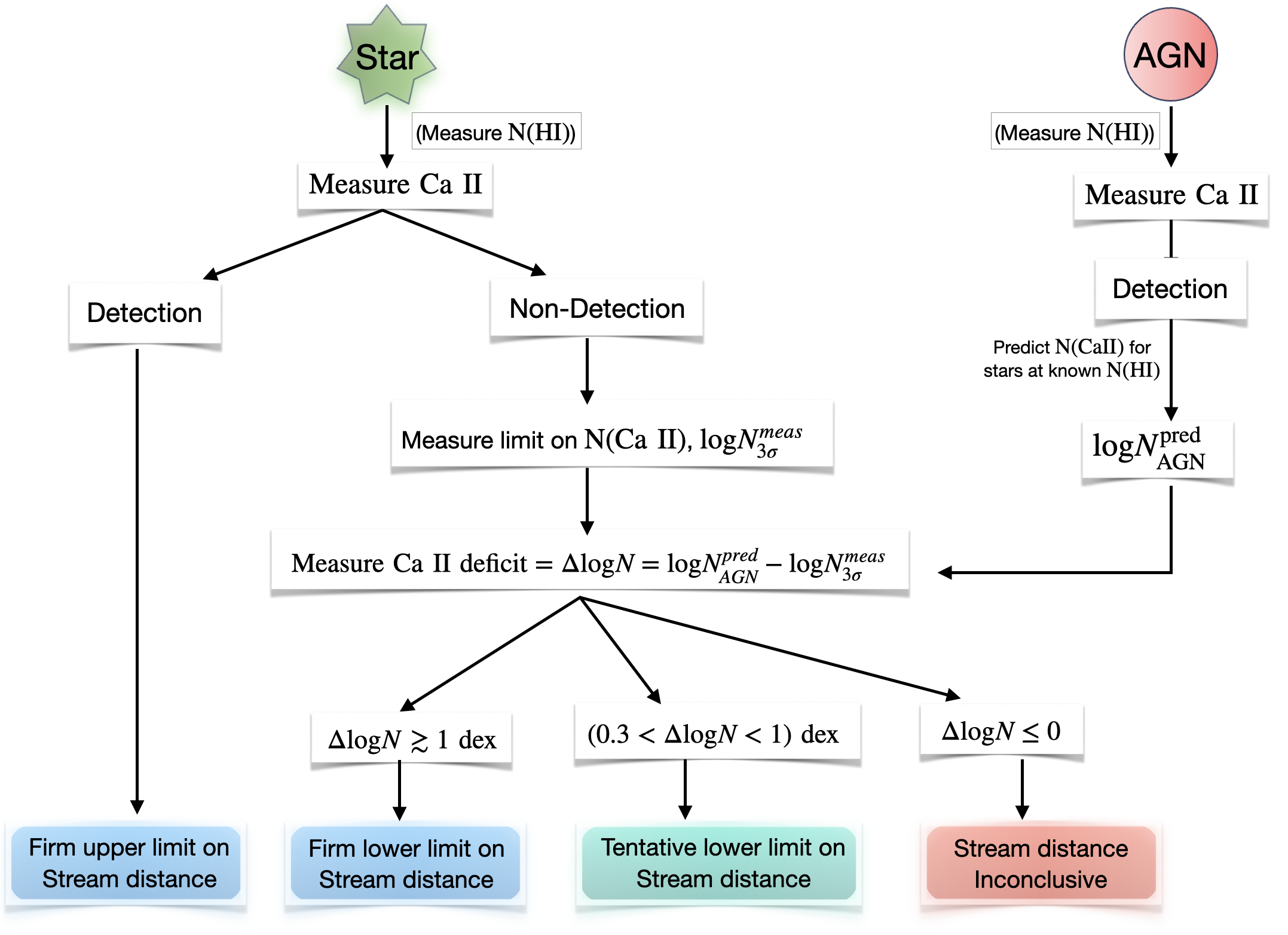}
 \caption{Flowchart illustrating the methodology used to constrain the Stream distance.} 
 \label{fig:method} 
\end{figure*}
%%%--------------
%%%%------------- Fig. 3 (Normalized CaII 3934 profiles) -----------
\begin{figure*}[!ht]
 \includegraphics[width=1.00\textwidth]{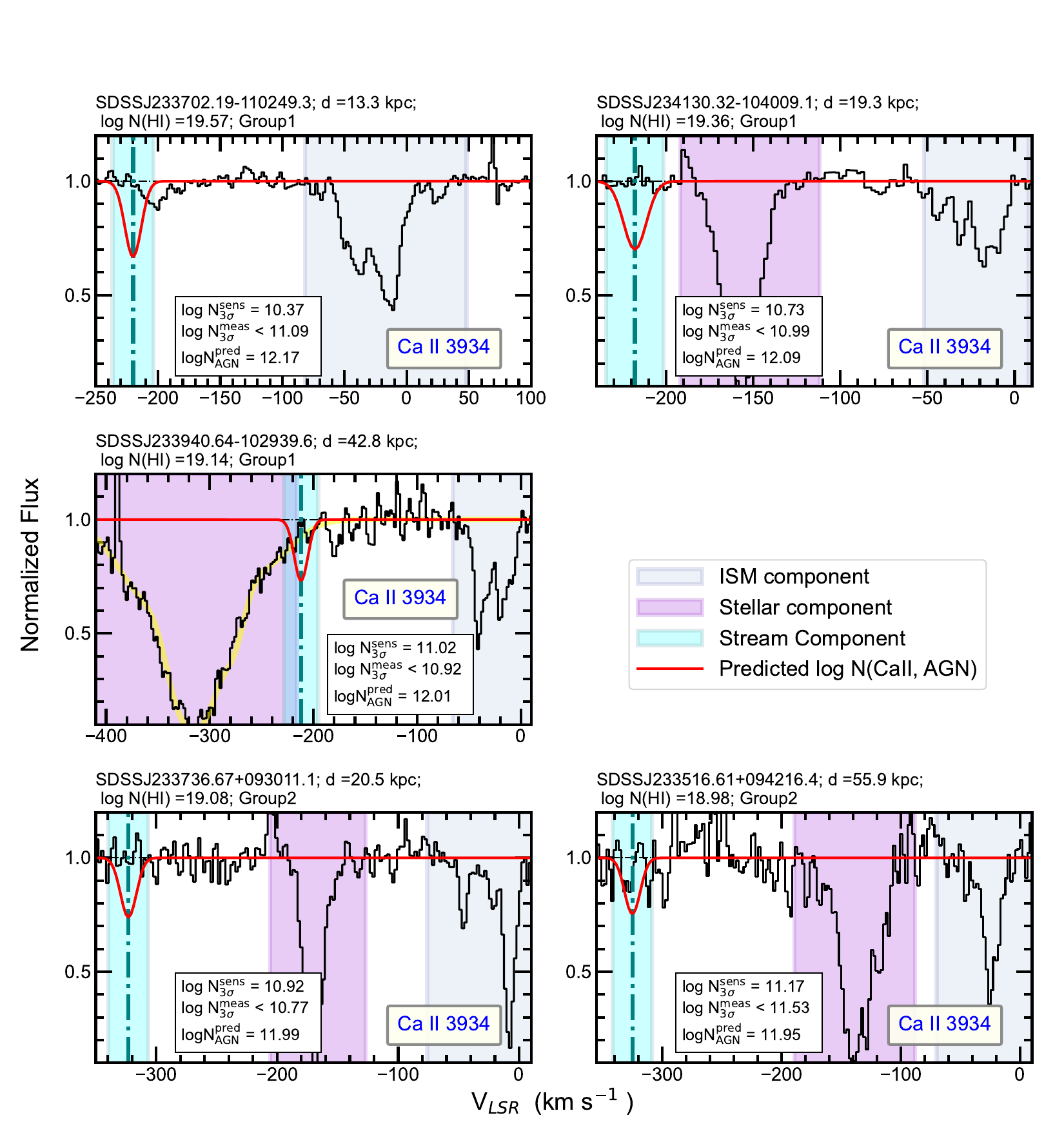}
 \caption{Normalized absorption profiles of \CaII\ 3934 as a function of LSR velocity. Each panel includes the name and distance of the star, the \HI\ column density of the Stream in proximity to the star, and the star group. The absorption components attributed to the stellar atmosphere (magenta), the ISM (steel blue), and the MS (aqua blue) are shaded within each panel. The LSR velocity corresponding to the MS is indicated by a dotted-dashed vertical line in teal \citep[based on the \HI\ emission map from][]{Westmeier2018}. The widths of the shaded regions for stellar and ISM absorption have been determined through visual inspection, while the shaded region representing Stream absorption is defined with a width of 16.5 \kms, which is three times the spectral resolution of the UVES Spectrograph (5.5 \kms). The predicted Gaussian profile for \CaII\ 3934, based on the expected column density $N$(\CaII) at observed $N$(\HI) predicted from the AGN studies, is plotted in a solid red curve in each panel, with an average velocity dispersion of $\sigma_{\rm Ca II} = $2\kms \citep[][]{Richter2013}. Each panel also includes annotations for the 3$\sigma$ sensitivity, the measured 3$\sigma$ limit on the absorption, and the predicted column density of \CaII. In the middle panel for the Group 1 BHB star SDSSJ233940.64$-$102939.6, the Voigt profile fit to the stellar component at $v_{\rm LSR}= -$315\kms\ is shown in yellow. The measured \Nmeas\ for this star is determined by normalizing the observed spectrum by the stellar profile fit to exclude the stellar contribution.}
 \label{fig:all_CaII} 
\end{figure*}
%%%--------------
\subsection{VLT/UVES Observations}\label{subsec:vlt-uves}

We observed our sample of 5 BHB stars using the VLT/UVES under ESO program ID 111.24PS.001 (PI: A. Fox). The observations were conducted in Service Mode with the 390+580 setting, employing 2$\times$2 binning and a 0.8" slit. This configuration provides wavelength coverage from 3260--4450~\AA\ and 4760--6840~\AA, designed to cover \CaII\ 3934, 3969~\AA\ and \NaI\ 5891, 5897~\AA\ absorption from the Stream. The spectral resolution for this setup 
($R\approx55,000$) corresponds to a velocity resolution of FWHM$\approx$5.5 \kms, allowing us to clearly distinguish the Stream absorption components from MS interstellar medium (ISM) and stellar components. The exposure times for each target are provided in Column 10 of Table~\ref{tab:sample_properties}. \par

We reduced the data using the standard UVES pipeline \citep{Ballester2000}
within the Common Pipeline Library environment, with calibration frames obtained close in time to the corresponding science frames. We subsequently performed preliminary pre-processing steps, including barycentric correction and spectral co-adding, to generate the final spectra in the heliocentric frame for all the targets. All wavelengths and velocities presented in this paper are then expressed relative to the Local Standard of Rest (LSR).\par

To analyze the \CaII\ and \NaI\ absorption features from the Stream in each spectrum, we focus on a spectral region spanning $\pm$1000 \kms\ around 
$v_{\rm LSR}$ = 0 \kms\ of the \CaII\ 3934, 3969~\AA\ and \NaI\ 5891, 5897~\AA\ transitions. Next, we employ the methodology outlined in \citet{Mishra2022} to accurately determine the local continuum level around these transitions in each sightline. In brief, each spectrum is initially smoothed over 51 pixels to derive a pseudo-continuum level, which is used to normalize the spectral region. To eliminate absorption features, an iterative boxed sigma-clipping process is applied to the pseudo-continuum-normalized spectra. The number of boxes is tailored to account for the presence or absence of prominent emission or absorption features, while the sigma thresholds are determined based on the median signal-to-noise ratio (SNR) of each spectrum. This iterative process is repeated until the residual spectrum is largely free of absorption features. The clipped spectral regions are subsequently interpolated linearly, and a spline is fitted to the interpolated spectrum to produce the final continuum-normalized spectrum for each stellar sightline.\par

Figures \ref{fig:stack_set1}, \ref{fig:stack_set2}, and \ref{fig:stack_set3} present the VLT/UVES spectra covering the \CaII\ 3934, 3969~\AA\ and \NaI\ 5891, 5897~\AA\ lines for our sample of five BHB stars. The absorption components originating from the stellar atmosphere, the MW ISM, and the Stream are marked in each panel. The spectral regions for stellar and MW ISM absorption are selected visually, while the spectral window for Stream absorption corresponds to a width of 16.5 \kms, which is three times the spectral resolution of UVES (5.5 \kms). %The red dashed curve in each panel represents the best-fit continuum level.
\par

\section{Constraining the Stream Distance} 
\label{sec:method}

\subsection{Distance Method}
\label{subsec:method}

To constrain the distance to the Stream, we adopted a methodology similar to that described in \citet[][]{Schwarz1995} and \citet[][]{Wakker1998} for the high- and intermediate- velocity clouds (HVCs and IVCs). 
This method involves analyzing the presence or absence of interstellar absorption lines such as \CaII\ and \NaI\ at the Stream's velocity in the spectra of stars with known distances.

The method involves the comparison of these absorption features at the Stream velocity in the optical spectra of stellar and background AGN sightlines. If absorption at the Stream's velocity is detected in the stellar spectra, the Stream must lie in front of the star. Conversely, the absence of such absorption indicates one of three possibilities: (1) the Stream is situated behind the star, (2) the line-of-sight traverses a low-density region of the Stream, or (3) the abundance of the relevant element in the Stream is too low for detection. To investigate scenarios (2) and (3), observations of AGNs positioned behind the Stream are especially informative. If absorption is detected toward an AGN with a known \HI\ column density at the Stream's location, it enables the determination of the \CaII/\HI\ (or \NaI/\HI) ratio along that line of sight. Assuming this ratio remains approximately uniform across the Stream, a comparable \HI\ column density measurement in the direction of a star can predict the expected \CaII\ (or \NaI) absorption strength, if the Stream lies in front of the star. It is also essential that the spectral sensitivity of the stellar observations is sufficient to detect the predicted absorption. If no \CaII\ absorption is observed, and the observed line strength is significantly weaker than the predicted line strength it can be concluded that the Stream is located beyond the star.\par

Note that the front-and-behind star experiment used here to constrain the Stream's distance assumes the Stream is either completely in front or completely behind each star, so it does not account for the possibility that the Stream lies \emph{partly} in front of a given star, and partly behind, given its finite thickness. If the entire Stream lies entirely behind the farthest star in our sample, this complication will not matter.

%%%%%%------------------ Table 3  (measurements) -------------
\begin{deluxetable*}{llllllllllllllll}
\label{tab:summary_stars}
\tabletypesize{\normalfont}
\tablewidth{0pt}\tabcolsep=3pt
\tablecaption{MS Absorption Measurements along the Stellar Sightlines}
\tablehead{Star Name                 &   $d$     &  log\,$N$\,(\HI)    &  Ion      &    S/N   &  \Wsens &  \Wmeas  &  \Nsens           & \Nmeas            &  \NpredAGN      & \dellogN  &  Result   \\
                                     &  (kpc)  &  (cm$^{-2}$)      &           &          &  m\AA   &  m\AA    &  (cm$^{-2}$)      &   (cm$^{-2}$)     &  (cm$^{-2}$)    &     (dex)       &          \\ 
           (1)			             &   (2)   &     (3)           &    (4)    &   (5)    &  (6)    &  (7)     &   (8)             &    (9)            &   (10)          &     (11)        & (12)     }
\startdata
SDSSJ233702.19-110249.3  &  13.3  &  19.57  &  Ca II~3934  &  61  &  2   &  9 $\pm$ 1    &  10.37  &  $<$11.09  &  12.17  & $>$1.08 & MND \\ 
                         &        &         &  Ca II~3969  &  50  &  3   &  10 $\pm$ 1   &  10.76  &  $<$11.45  &    ...  &  ... & ... \\
                         &        &         &  Na I~5891   &  50  &  4   &  4 $\pm$ 1    &  10.28  &  $<$10.58  &    ...  &  ... & ... \\
                         &        &         &  Na I~5897   &  55  &  3   &  4 $\pm$ 1    &  10.53  &  $<$10.85  &    ...  &  ... & ... \\\\
SDSSJ234130.32-104009.1  &  19.3  &  19.36  &  Ca II~3934  &  27  &  5   &  4 $\pm$ 2    &  10.73  &  $<$10.99  &  12.09  & $>$1.1  & MND \\ 
                         &        &         &  Ca II~3969  &  27  &  5   &  7 $\pm$ 2    &  11.03  &  $<$11.41  &    ...  &  ... & ... \\
                         &        &         &  Na I~5891   &  31  &  6   &  1 $\pm$ 2    &  10.48  &  $<$10.52  &    ...  &  ... & ... \\
                         &        &         &  Na I~5897   &  33  &  6   &  -7 $\pm$ 2   &  10.75  &  ...    &    ...  &  ... & ... \\\\
SDSSJ233940.64-102939.6  &  42.8  &  19.14  &  Ca II~3934  &  14  &  9   &  -2 $\pm$ 3   &  11.02  &  $<$10.92  &  12.01  & $>$1.09 & MND \\ 
                         &        &         &  Ca II~3969  &  24  &  5   &  60 $\pm$ 2   &  11.08  &  $<$12.18  &    ...  &  ... & ... \\
                         &        &         &  Na I~5891   &  27  &  7   &  -6 $\pm$ 2   &  10.54  &  $<$9.37   &    ...  &  ... & ... \\
                         &        &         &  Na I~5897   &  28  &  7   &  10 $\pm$ 2   &  10.83  &  $<$11.24  &    ...  &  ... & ... \\\\
SDSSJ233736.67+093011.1  &  20.5  &  19.08  &  Ca II~3934  &  17  &  7   &  -2 $\pm$ 2   &  10.92  &  $<$10.77  &  11.99  & $>$1.22 & MND \\ 
                         &        &         &  Ca II~3969  &  15  &  8   &  14 $\pm$ 3   &  11.28  &  $<$11.72  &   ...   &  ... & ... \\
                         &        &         &  Na I~5891   &  29  &  6   &  -4 $\pm$ 2   &  10.51  &  $<$10.12  &   ...   &  ... & ... \\
                         &        &         &  Na I~5897   &  25  &  8   &  163 $\pm$ 3  &  10.88  &  $<$12.24  &   ...   &  ... & ... \\\\
SDSSJ233516.61+094216.4  &  55.9  &  18.98  &  Ca II~3934  &  10  &  13  &  17 $\pm$ 4   &  11.17  &  $<$11.53  &  11.95  & $>$0.42 & TND \\ 
                         &        &         &  Ca II~3969  &  9   &  14  &  58 $\pm$ 5   &  11.5   &  $<$12.22  &  ...    &  ... & ... \\
                         &        &         &  Na I~5891   &  13  &  15  &  2 $\pm$ 5    &  10.87  &  $<$10.92  &  ...    &  ... & ... \\
                         &        &         &  Na I~5897   &  13  &  14  &  302 $\pm$ 5  &  11.17  &  $<$12.51  &  ...    &  ... & ... \\\\
Stack1  &  25$\pm$13  &  19.39  &  Ca II~3934  &  72  &  3  &  5 $\pm$ 1 & 10.54  &  $<$10.99  &  12.1 & $>$1.11   & MND \\
        &             &         &  Na I~5891  &  104  &  3  &  2 $\pm$ 1 & 10.2   &  $<$10.39  &  ...  &  ...   & ... \\\\
Stack2  &  38$\pm$1   &  19.03  &  Ca II~3934  &  37  &  6  &  -1 $\pm$ 2 & 10.83  &  $<$10.78  &  11.97 & $>$1.19 & MND \\
        &             &         &  Na I~5891   &  38  &  8  &  0 $\pm$ 3  & 10.63  &  $<$10.64  &  ...  &  ...  & ... \\\\
\enddata
\tablecomments{Col. (1) Star name; Col. (2) Distance from the Sun; Col. (3) \HI\ column density of the Stream near the star location; Col. (4) name of the ion; Col. (5) Median SNR per pixel of the line free region near the Stream velocity; Col. (6) 3$\sigma$ equivalent width sensitivity; Col. (7) measured equivalent width at the Stream velocity integrated over a velocity interval three times the UVES resolution element;
Col. (8) 3$\sigma$ column density sensitivity, calculated from column (6) assuming a linear curve-of-growth;
Col. (9) 3$\sigma$ maximum allowed column density, calculated from column (7) assuming a linear curve-of-growth, and including the 3$\sigma$ error;
Col. (10) Predicted column density from AGN sightlines assuming linear relationship between $N$(\HI) and $N$(\CaII) values in Table~\ref{tab:agn_summary}. 
Col (11) \caldef\ defined as the difference in the \NpredAGN\ and \Nmeas\ values; Col (12) Result: meaningful non-detection (MND) or tentative non-detection (TND). A MND is obtained when \dellogN $\ga$1.0\,dex , while TND is when  0.3 $<$ \dellogN $<$1.0\,dex (see Fig.~\ref{fig:method}).}
\end{deluxetable*}
%%%%%%%%%%%%%-----------------------------------

\subsection{Measuring Limits on Absorption} \label{subsec:measurements}

We use two methods to measure the strength of \CaII\ and \NaI\ absorption at the Stream velocity in each of our five stellar sightlines: (1) calculating the spectral sensitivity, and (2) measuring the \emph{actual} absorption-line equivalent width at the Stream velocity. The first quantity, the 3$\sigma$ spectral sensitivity, is denoted as \Wsens\ and calculated for an unresolved line using: \( \text{$W$}^{\rm sens}_{3\sigma} = 3 \text{(SNR)}^{\rm -1} \Delta\lambda_{\rm instr} \)~\AA, where SNR is computed over a line-free region within a resolution element $\Delta \lambda_{\rm instr}$. For VLT/UVES $\Delta \lambda_{\rm instr}$ is defined from the relation R = $\lambda/\Delta \lambda_{\rm instr} \approx$ 55,000 at the wavelength of interest.
%\( \text{$W$}^{\rm sens}_{3\sigma} = 3 \text{(SNR)}^{\rm -1} (\lambda^{'} / \text{$R$}) \)~\AA; where SNR is computed over a line-free region within a width of $\pm10$ \AA\ around the wavelength $\lambda^{\prime}$ (see Column 5 of Table~\ref{tab:summary_stars}) and $R$ is the resolution of the UVES spectrograph ($R\approx 55,000$). 
This quantity is the \emph{theoretical} limit on the minimum detectable absorption feature, because it is derived from a completely line-free region. The second quantity, the measured equivalent width (\Wmeas), is determined by integrating the spectrum over a region spanning three times the UVES resolution around the Stream velocity. This quantity is the \emph{observational} limit on the absorption. It can differ from the theoretical limit because of the presence of absorption in the data, either real or contaminating (for example data reduction artifacts).
The values of \Wsens\ and \Wmeas\ are listed in Columns 6 and 8 of Table~\ref{tab:summary_stars}, respectively. We convert \Wsens\ to column density in Column (8) of Table~\ref{tab:summary_stars}, assuming the lines lie on the linear part of the curve-of-growth (COG), using the equation: 

\[N =  1.13 \times 10^{20} \, \frac{\text{$W$}}{f \lambda^{2}} \, \text{cm}^{-2}  ~~.\] 

where $\lambda$ is the wavelength in \AA\ and $f$ is the oscillator strength of the line under consideration. The value in Column (9) of Table~\ref{tab:summary_stars} represents the maximum allowed column density at 3$\sigma$. It is calculated by adding the 3$\sigma$ uncertainty to the \Wmeas\ values and applying a linear COG equation.\par

In the case of a non-detection, we need to confirm if the non-detection is meaningful, 
i.e. less than the expected line strength,
so we can establish a firm lower limit on the Stream distance along this sightline. To do this, we compare our $3\sigma$ measured column densities (\Nmeas) with the measured column densities along Stream sightlines to AGNs \citep{Fox2010,Fox2013,Richter2013}, obtained using optical VLT/UVES spectra. These studies provide five AGN sightlines with comparable \HI\ columns to our sightlines with available \CaII\ and \NaI\ absorption line data (see Fig.~\ref{fig:sample}). The \CaII\ and \NaI\ results for these sightlines are summarized in Table~\ref{tab:agn_summary}. Using Columns (2) and (4) of Table~\ref{tab:agn_summary}, and assuming a linear relationship between $N$(\HI) and $N$(\CaII), we perform a linear fit to establish the relationship between these two quantities. The resulting relation is given by: \( \log N^{\rm pred}_{\rm AGN}(\mathrm{CaII}) = 0.37 \times \log N(\mathrm{H\,I}) + 4.93,\) where we include only AGN sightlines with \logNHI~$>$18.5 to ensure consistency with our stellar sightlines. 

Using this relation and the measured $N$(\HI) in each sightline, we derive the predicted \CaII\ column density (\NpredAGN) along each stellar sightline is listed in Column (10) of Table~\ref{tab:summary_stars}. Notably, a linear relationship could only be derived for the \CaII\ 3934 line. For \NaI\ 5891,5897~\AA, Stream absorption is detected along only one AGN sightline, resulting in a highly uncertain predicted $N$(\NaI) column density toward the stars. Additionally, the \CaII\ 3969~\AA\ Stream region in our stellar spectra is blended with the stellar Balmer epsilon line at 3970~\AA. Therefore, we refrain from drawing any conclusions from the non-detection of \NaI\ 5891,5897~\AA\ or of \CaII\ 3969~\AA.
\par

We caution that the assumption of a linear relationship between $N$(\HI) and $N$(\CaII) may not always hold. Variations in ionization conditions, UV radiation fields, dust depletion, line saturation, or small-scale structures within the Stream could introduce deviations from linearity, which would affect the predicted \CaII\ column densities. Consequently, we only flag stellar sightlines as providing a meaningful non-detection (MND) when \Nsens\ is lower than \NpredAGN, and when the difference \dellogN\ = \NpredAGN -- \Nmeas\ is more than 1.0 dex. We refer to this difference as the \caldef. This provides a safety factor to account for the aforementioned physical effects.
\citep[see][for the discussion on safety factor used for HVCs and IVCs]{Wakker2001,Vanloon2009,Smoker2011, Smoker2015}.
In Fig.~\ref{fig:method}, we present a flowchart illustrating step-by-step our method for constraining the distance of the Stream.

\subsection{Stacking of Stellar Spectra} 
\label{subsec:stack_measurements}

To further enhance the spectral sensitivity for detecting weak \CaII\ and \NaI\ absorption, 
we stack the stellar spectra into two subgroups, as defined later in this section. We employ the SNR-weighted spectral stacking technique described in \citet{Mishra2024a}. To summarize the stacking method, for each stacking group, we first shift the spectra of all contributing stellar sightlines to the Stream's rest frame by dividing the observed wavelengths by $(1 + v_{\rm Stream}/c)$, where $v_{\rm Stream}$ is given in Column 8 of Table~\ref{tab:sample_properties} and $c$ is the speed of light. Next, we collect the normalized raw flux values from these stars within a velocity window of $\pm$500~\kms\ in the Stream's rest frame, centered around the given line (i.e., \CaII\ or \NaI). This velocity window is divided into bins of 6~\kms, approximately matching the spectral resolution of UVES ($\Delta v \approx 5.5$\kms). For each velocity bin, we calculate the mean flux weighted by the continuum SNR near the line of interest (see Column 5 of Table~\ref{tab:summary_stars}) to generate the final stacked profiles for the \CaII\ and \NaI\ lines in the Stream's rest frame. The SNR-weighted mean is preferred because it minimizes sensitivity to variations in the SNR of individual contributing spectra, as demonstrated in \citet{Mishra2024a}.\par

\begin{figure}[!hbt]
\hspace{-0.2in}
 \includegraphics[width=0.5\textwidth]{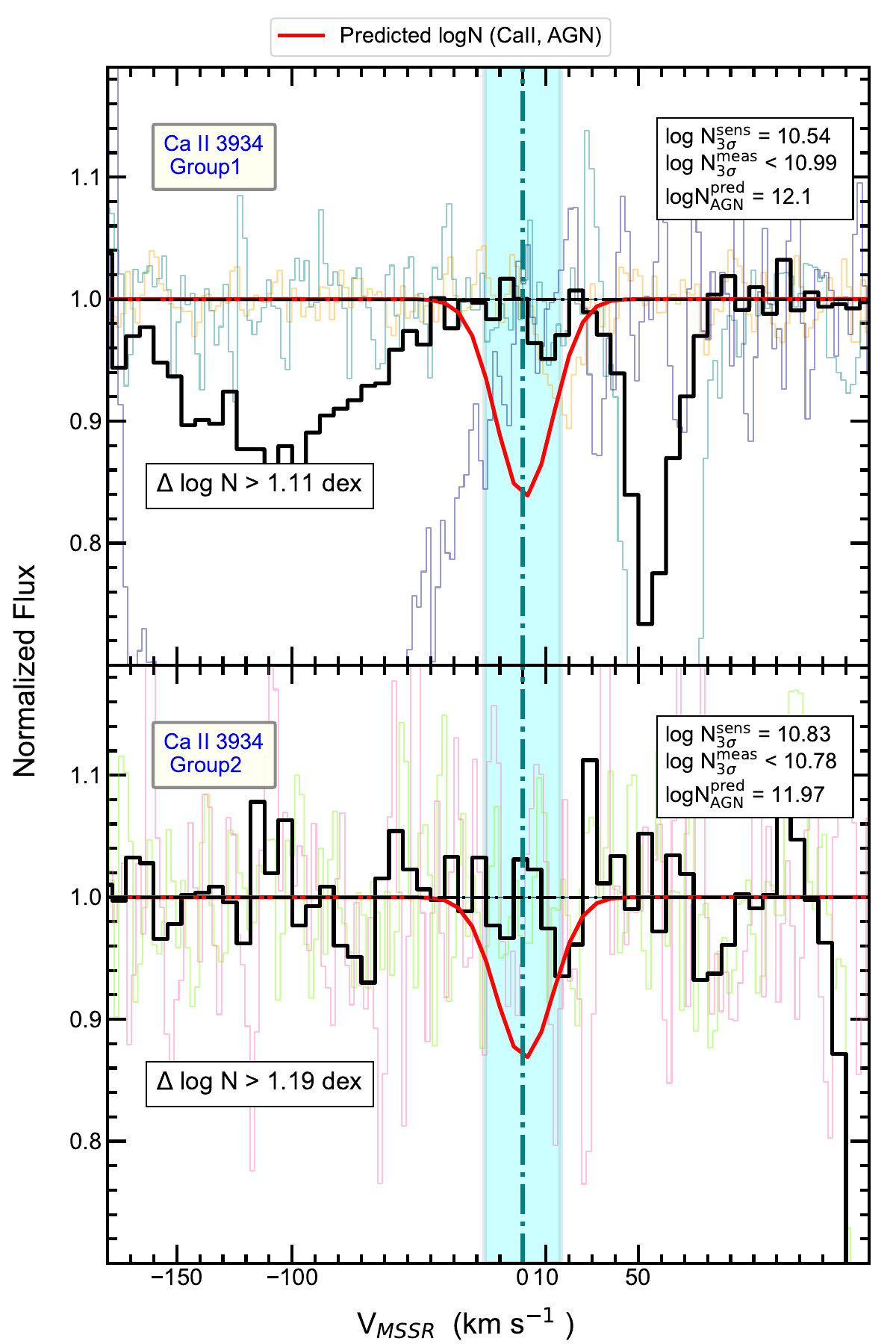}
 \caption{SNR-weighted stacked \CaII\ 3934 profiles in the rest frame of the Stream (so $v_{\rm MSSR} = 0$ \kms\ corresponds to the Stream velocity). The top panel shows the stacked \CaII\ profiles for Group 1 stars and the bottom  panel for Group 2 stars. The individual stellar spectra contributing to each stack are also shown in each panel. The 3$\sigma$ sensitivity limits and the measured 3$\sigma$ absorption strengths derived from the stacked profiles are indicated in each panel. The predicted \CaII~3934 column density derived from AGN sightlines (\NpredAGN), based on the average $N$(\HI) is represented by a solid red curve in each panel.}
 \label{fig:wmean_stacks_CaII} 
\end{figure}

Following this procedure, we generate two stacked profiles for both the \CaII\ and \NaI\ lines: (1) the Group 1 stack, combining spectra from the three Group 1 stars; (2) the Group 2 stack, combining spectra from the two Group 2 stars. The stacked profiles for the \CaII~3934 line are presented in Fig.~\ref{fig:wmean_stacks_CaII} for Group 1 in the top panel and for Group 2 in the bottom panel. The corresponding stacked profiles for the \NaI~5891 line are shown in Appendix Fig.~\ref{fig:wmean_stacks_NaI}. The measurements derived from these stacked profiles for both \CaII~3934 and \NaI~5891 are summarized in Table~\ref{tab:summary_stars}. To calculate \NpredAGN\ for the stacked profiles, we use N(\HI) values calculated from the mean of the N(\HI) values from the contributing sightlines in the stack. This approach is justified as the stellar sightlines probe a smaller region of the Stream, where $N$(\HI) does not vary substantially. The distances for the stacked profiles are defined as the mean of the stellar distances contributing to the stacks, with a 1$\sigma$ scatter representing the standard deviation of these distances.\par

%%%%------------- Fig. 4 (CaII stack)-----------
%%%--------------

\section{Results}
\label{sec:results}

\subsection{Results for Group 1}
\label{subsec:res_group1}

As illustrated in Fig.~\ref{fig:sample}, three of our stars are located in the region classified as Group 1. The logarithmic \HI\ column density of the Stream in these
sightlines are 19.14, 19.36 and 19.57, with velocities of $-$212\kms, $-$218\kms, and $-$220\kms\ respectively (average $-$217\kms). This region lies approximately 80 degrees from the Magellanic Clouds. The distances to the three stars are 13.3 kpc, 19.3 kpc, and 42.8 kpc respectively from the Sun (see Table~\ref{tab:sample_properties}). All three sightlines exhibit non-detection of \CaII\ and \NaI\ absorption lines at the Stream velocities, as shown in Figs.~\ref{fig:stack_set1} and \ref{fig:stack_set2}.\par

To explore whether the non-detection of Stream absorption along these sightlines is meaningful, we compare the limit on the measured \CaII\ column density (\Nmeas) with the predicted \CaII\ column density (\NpredAGN). Since \NpredAGN\ can only be derived for the \CaII~3934 line, we concentrate exclusively on this line. The normalized \CaII~3934 line profiles and measurements are presented in the top three panels of Fig.~\ref{fig:all_CaII} for Group 1 stars, with measurement limits summarized in Table~\ref{tab:summary_stars}. The spectral sensitivity at \CaII~3934 (\Nsens) is sufficient to detect \NpredAGN\ for all the three sightlines in Group 1. A comparison between \Nmeas\ and \NpredAGN\ reveals that these sightlines show a \caldef\ (\dellogN) $>$1.0 dex (see Column 11 of Table~\ref{tab:summary_stars}), confirming the non-detections are meaningful. The distances for these sightlines are 13.3 kpc, 19.3 kpc, and 42.8 respectively (see Table~\ref{tab:sample_properties}), leading us to conclude that toward Group 1 the Stream must lie beyond 42 kpc from us.

%two of the Group 1 sightlines show a \caldef\ (\dellogN) $>$1.0 dex,  confirming the non-detections are meaningful. The distances for these two sightlines are 13.3 kpc and 19.3 kpc, respectively, leading us to conclude that toward Group 1 the Stream must lie beyond 19.3 kpc from us. The third sightline, at a distance of 42.8 kpc, shows \dellogN$>$0.42 dex, which we consider a tentative non-detection. Therefore, the Stream is tentatively constrained to be at $d>42.8$ kpc towards Group 1,  but this needs confirmation from deeper spectra.\par

\subsection{Results for Group 2}
\label{subsec:res_group2}

The Group 2 region includes two BHB stars at distances of 55.9 kpc and 20.5 kpc (see Fig.~\ref{fig:sample}). This region has higher Magellanic longitude than Group 1, by $\approx$20$\deg$, and is therefore further from the LMC and SMC. The logarithmic \HI\ column densities are 18.98 and 19.08 along these two stellar sightlines, with corresponding Stream velocities of $-$325 \kms\ and $-$323 \kms.\par

No detection of \CaII\ and \NaI\ absorption at the Stream velocities is seen in any of the Group 2 directions (see Fig.~\ref{fig:all_CaII} and Fig.~\ref{fig:stack_set3}. The spectral sensitivity at \CaII~3934 is sufficient to detect \NpredAGN\ in both Group 2 sightlines. For the 20.5 kpc sightline, \dellogN$>$1.0 dex, indicating a robust non-detection that implies the Stream in this region is at least 20.5 kpc away. However, for the more distant sightline at 55.9 kpc, \dellogN$>$0.41 dex. This less-firm non-detection suggests that the Stream is at $d>55.9$ kpc, but this is not a conclusive limit given the data quality and hence needs confirmation from deeper spectra.\par

%In summary, our analysis confidently constrains the lower limit of the Stream’s distance to be $>$20 kpc based on stars from both groups. Additionally, we infer a more tentative lower limit of 42 kpc for the Group 1 region and 55 kpc for the Group 2 region. These constraints are derived from two distinct regions of the Stream, accounting for the possibility that the Stream distance depends on position.\par

\subsection{Results from Stacking}\label{subsec:res_stack}

The purpose of the stacking analysis is to enhance the spectral sensitivity, enabling the detection of very weak signals that may only become apparent after stacking. 
As described in Section~\ref{subsec:stack_measurements}, we apply an SNR-weighted method to stack the three stellar sightlines from Group 1, and the two stellar sightlines 
from Group 2. These stacks allow us to extend the lower limit on the average distance of the Stream toward each group, under the assumption that the Stream lies behind the farthest star in the stacked profile. This assumption is reasonable, given that no absorption is detected from the farthest stars in both groups.

As evident in Fig.~\ref{fig:wmean_stacks_CaII} and in Table~\ref{tab:summary_stars}, the \CaII\ deficit (\dellogN) measured in both the Group 1 and Group 2 stacks is $>$1.0 dex, so that the stacks give meaningful non-detections, which lead to robust lower limits on the Stream's average distance of 25$\pm$13 kpc and 38$\pm$17 kpc for Group 1 and Group 2, respectively. These distances are calculated as the averages of the distances of the member stars in Group 1 and Group 2. \par

\section{Discussion}
\label{section:discussion}

The distance to the Stream has been a long-standing issue. Early simulations of the formation of the Stream in a first infall scenario placed the Stream at distances ranging from 100 to 200 kpc \citep{Besla2012, Pardy2018, Lucchini2020}. However, a more recent simulation by \citet{Lucchini2021}, incorporating first-passage orbit and the effects of both the MW and Magellanic coronae, suggests that the closest point of the Stream is approximately 20 kpc from the Sun. However, until now, there have been no observational constraints on the distance to the gaseous component of the Magellanic Stream.\par

%%%%------------- Fig. 4 (result summary)-----------
\begin{figure}[!ht]
%\hspace{-0.3in}
 \includegraphics[width=0.45\textwidth]{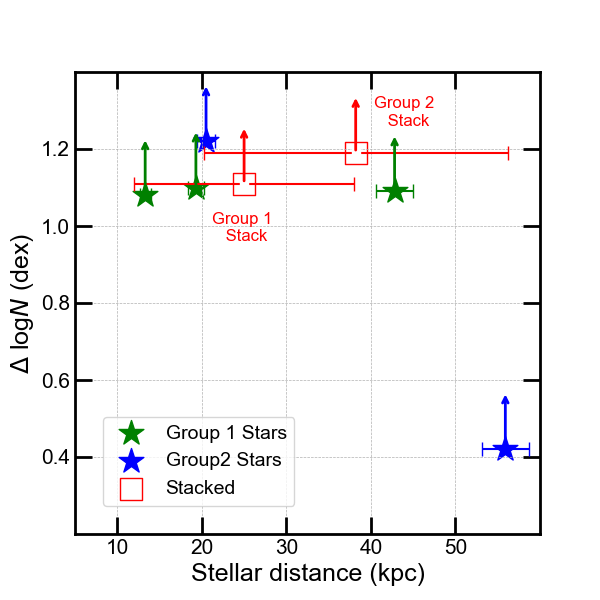}
 \caption{The \caldef\ (\dellogN), defined as the difference between the predicted column density (\NpredAGN) and the measured column density (\Nmeas) for the \(\mathrm{Ca\,II}\,3934\) line, is plotted against stellar distances. Individual sightlines for Group 1 and Group 2 stars are represented by green and blue stars, respectively, while \(\Delta \log N\) values for stacked profiles are shown as open red squares. Upward arrows indicate that these data points represent lower limits of \dellogN. The errors on the stellar distances 
are 5\% of their respective values \citep[see][]{Xue2011}.}
 \label{fig:res_summary} 
\end{figure}
%%%--------------

Our study provides the first observational constraints on the distance to the gaseous Stream. Figure~\ref{fig:res_summary} summarizes the findings from individual sightlines and stacked profiles. The figure presents \dellogN\ (the \CaII\ deficit) as a function of stellar distance for both individual sightlines (green and blue stars) and stacked profiles (open red squares). A positive \CaII\ deficit yields a lower limit on the distance to the Stream. All sightlines in the Group 1 region out to 42 kpc exhibit clear and meaningful non-detections with $\Delta \log N \gtrapprox 1.0$ dex. These 3$\sigma$ non-detections are unlikely to result from lower $N$(\HI) or reduced \CaII\ abundance in the Stream, providing a robust lower limit of the Stream’s distance to at least 42 kpc for this region. In Group 2, the sightline at 20.5 kpc also shows $\Delta \log N \gtrapprox 1.0$ dex, firmly placing the Stream beyond 20 kpc. Although the star at  55.9 kpc in this region also give \CaII\ non-detections, but with the \CaII\ deficit of only 0.4\,dex. So while this non-detections suggest that the Stream distances is $>$55.9 kpc towards Group 2, the result is not conclusive. The results from the stacked sightlines support a more distant Stream with \dellogN $\gtrsim$ 1.0 dex.\par

%The two stars at 42.8 (Group 1) and 55.9 kpc (Group 2) also give \CaII\ non-detections, but in these sightlines the \CaII\ deficit is only 0.4\,dex. So while these non-detections suggest that the Stream distances are $>$42.8 and $>$55.9 kpc in the two regions,  the results are not conclusive. The results from the stacked sightlines support a more distant Stream with \dellogN $\gtrsim$ 1.0 dex.

\subsection{The Stellar Stream}\label{subsec:stellar-stream}

\citet{Zaritsky2020} presented a sample of 15 H3 stars with galactocentric distances greater than 40 kpc, and with a similar sky position and velocity to the gaseous Magellanic Stream. The authors found that the chemical abundances of their stars are consistent with those of SMC stars, suggesting an SMC origin
\citep[see also][]{Zaritsky2024}. Later, \citet{Chandra2023} identified a stellar counterpart to the gaseous Stream, studying 13 stars with extreme angular momentum and distances beyond 50 kpc. These stars exhibit kinematics and past orbits similar to the Magellanic Clouds and show a metallicity bifurcation forming thin stellar Streams aligned with the \HI\ Stream, further supporting their connection to the Magellanic Stream.

Notably, some stars from \citet{Zaritsky2020} and \citet{Chandra2023} overlap with the two regions examined here. In this study, we establish a robust lower limit of 20 kpc and a tentative lower limit of 55 kpc for the gaseous Stream, consistent with the distances derived for the stellar component \citep{Zaritsky2020, Zaritsky2024, Chandra2023}. However, additional stellar sightlines beyond 50 kpc are needed to better constrain the location of the gaseous Stream and its connection to the stellar stream.\par

\section{Summary}
\label{sec:summary} 
We have analyzed the VLT/UVES spectra of five BHB stars tracing two regions, Group 1 (l$_{\rm MS}$, b$_{\rm MS}$ $\approx$ $-$79\degree, $-$2\degree) and Group 2 (l$_{\rm MS}$, b$_{\rm MS}$ $\approx$ $-$98\degree, 3\degree) near the tip of the trailing Stream. We have analyzed the \CaII\ and \NaI\ absorption at the Stream's velocity to derive constraints on the Stream's distance. No detectable \CaII\ or \NaI\ absorption was found at the Stream velocities across any of these five sightlines. By comparing the measured column densities of the \CaII~3934 with those determined from AGN sightlines, we define the \caldef\ (\dellogN) as a metric to quantify the meaningfulness of the non-detections. From sightlines with \dellogN\ $>$ 1 dex (meaningful non-detections), we establish robust lower limits of 42 kpc and 20 kpc on the Stream's distance for Group 1 and Group 2 regions respectively. With \dellogN\ $>$ 0.4 dex, we also place a tentative lower limit of 55 kpc for the Group 2 region. Additionally, by separately stacking the spectra of Group 1 and Group 2 stars using an SNR-weighted method, we determine lower limits on the average Stream distance of 25 kpc and 38 kpc, respectively, for the two groups.\par

Although this study relies entirely on non-detections, it provides crucial constraints on the MS distance, which has lacked observational limits until now. Additional stellar sightlines with high-sensitivity spectra at larger distances, particularly beyond 55 kpc, are necessary to further refine the distance constraints.

%%%--------------

%\begin{acknowledgments}
\vspace{0.2 cm}
{\it Acknowledgments:}
 We thank the referee Prof. David Nidever for the constructive comments on our manuscript. We thank the Milky Way Halo group at Space Telescope Science Institute for the useful discussion.

\facilities{VLT-UVES \citep{Dekker2000}, GASS \citep{McClure-Griffiths2009}}
\software{UVES pipeline \citep{Ballester2000}, \texttt{astropy} \citep{astropy2018}, 
\texttt{Magellanic Stream Nidever08} \citep{Nidever2008}.}

\appendix

\setcounter{figure}{0}
\renewcommand{\thefigure}{A\arabic{figure}}
\renewcommand{\thetable}{A\arabic{table}}  
\renewcommand{\thesection}{A\arabic{section}}
%\section{Appendix information}

%%%%%------------- App Data fig 1 -----------
\begin{figure*}[!hbt]
 \includegraphics[width=1\textwidth]{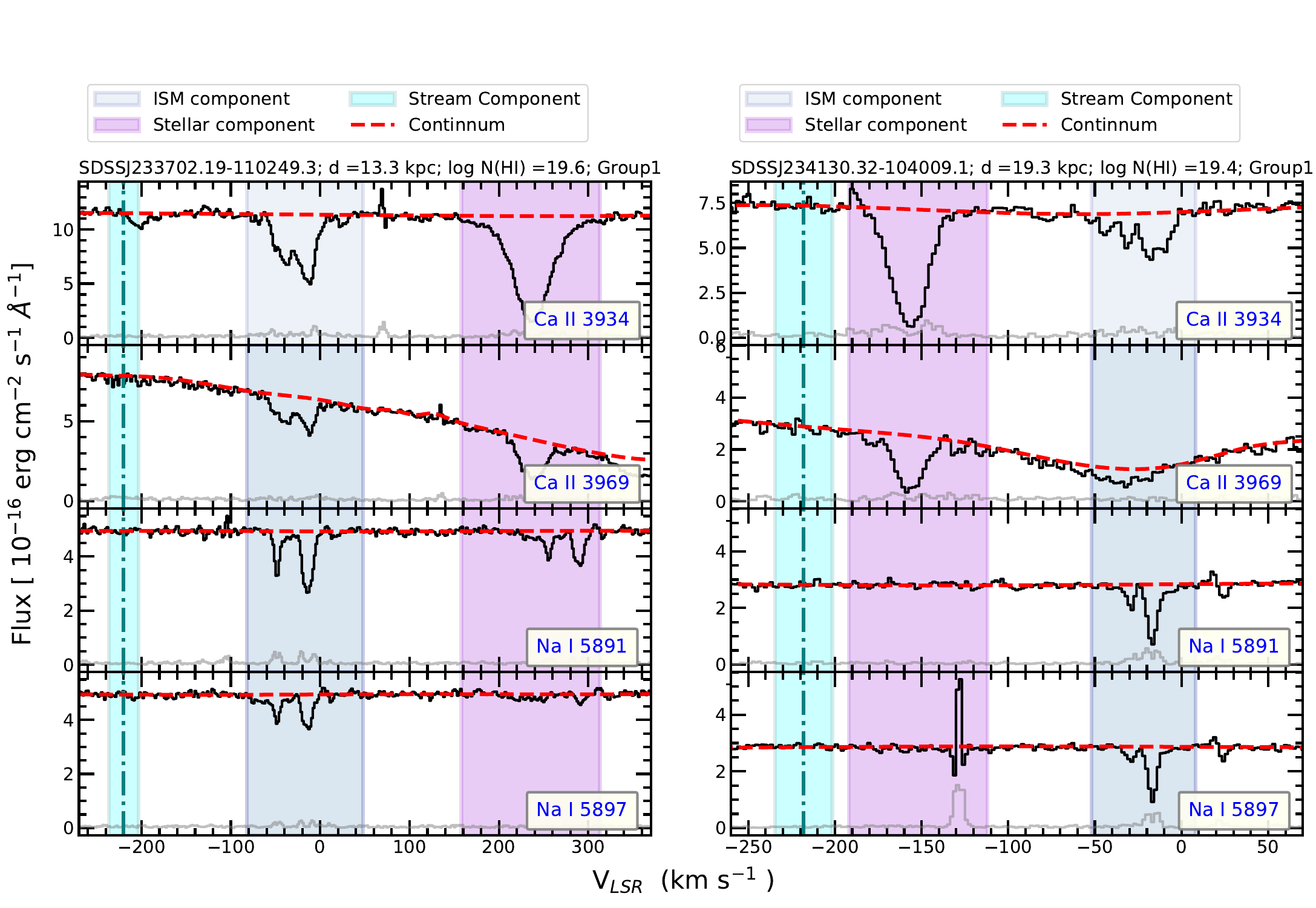}
 \caption{\emph{Left}: Unnormalized absorption profiles of \CaII\ and \NaI\ in the VLT/UVES spectrum of the Group 1 BHB star SDSS J233702.19$-$110249.3, plotted against the LSR velocity. The errors on the profiles are plotted in gray at the bottom of each panel. The name and distance of the star, the \HI\ column density of the MS near the star’s location, and the star group are provided at the top of each panel. The red dashed curve represents the best-fit continuum level. Absorption from the MS is expected at an LSR velocity of $\approx-$220 \kms\ \citep[based on the \HI\ emission map from][]{Westmeier2018}. The absorption components originating from the stellar atmosphere (magenta), the ISM (steel blue), and the MS (aqua blue) are shaded and labeled in each panel. The width of the shaded regions corresponding to stellar and ISM absorption are determined visually, while the shaded region for Stream absorption corresponds to a width of 16.5 \kms, which is three times the spectral resolution of UVES (5.5 \kms). Note that the \CaII\ 3969 absorption profile is contaminated by the stellar hydrogen Balmer line (H$\varepsilon~$3970 \AA). \emph{Right:} Same as \emph{left}, but for the Group 1 BHB star SDSS J234130.32$-$104009.1, with an expected Stream velocity of $\approx-$218 \kms \citep{Westmeier2018}.}
 \label{fig:stack_set1} 
\end{figure*}
%%%--------------

%%%%-------------App  Data fig 2 -----------
\begin{figure}[!hbt]
\hspace{-0.2in}
 \includegraphics[width=0.5\textwidth]{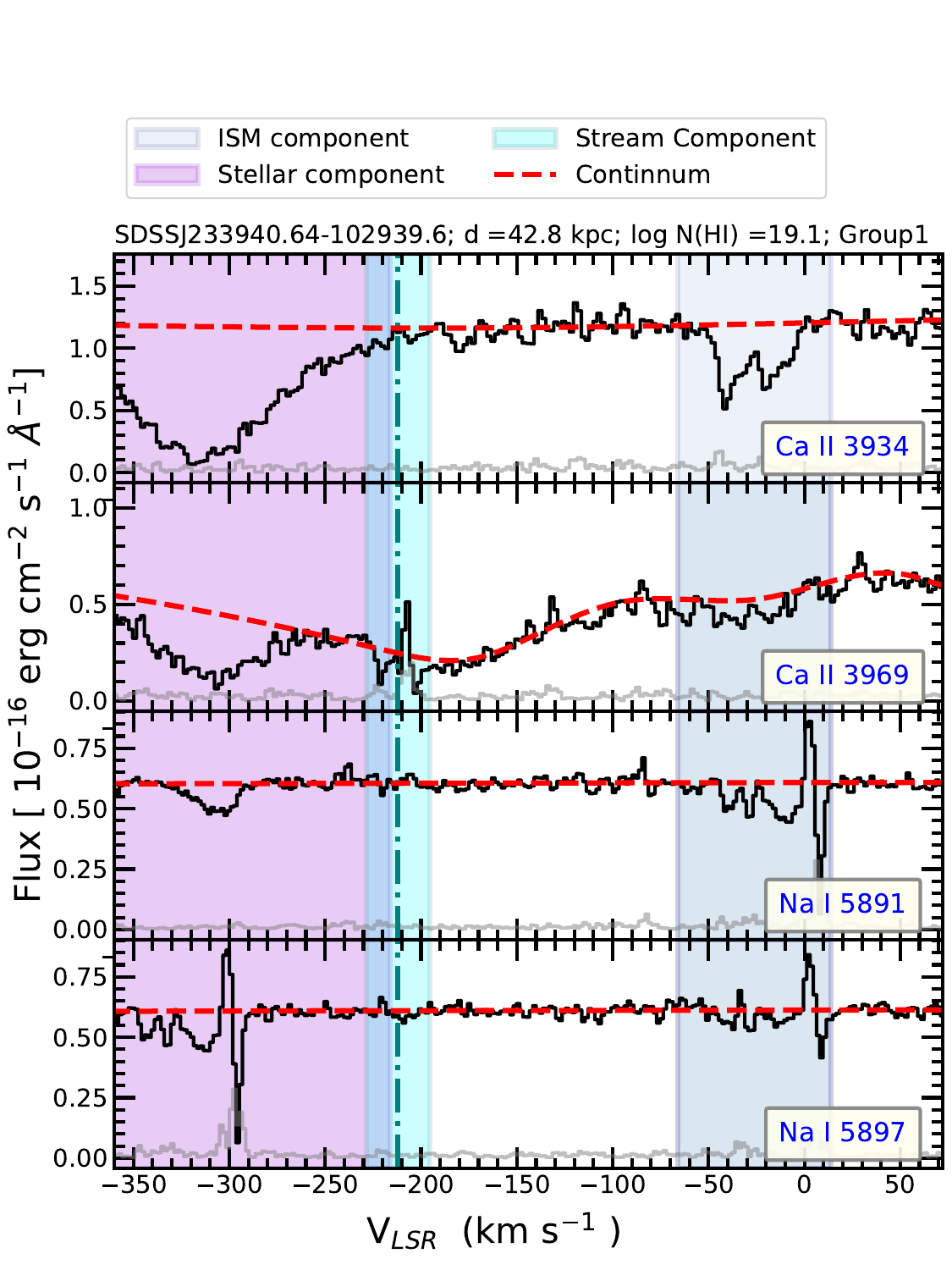}
 \caption{Same as Fig.~\ref{fig:stack_set1} but for the Group 1 BHB star SDSSJ233940.64$-$102939.6 with an expected Stream velocity of $\approx-$212 \kms.}
 \label{fig:stack_set2} 
\end{figure}
%%%--------------

%%%%------------- App Data fig 3 -----------
\begin{figure*}[!hbt]
 \includegraphics[width=1\textwidth]{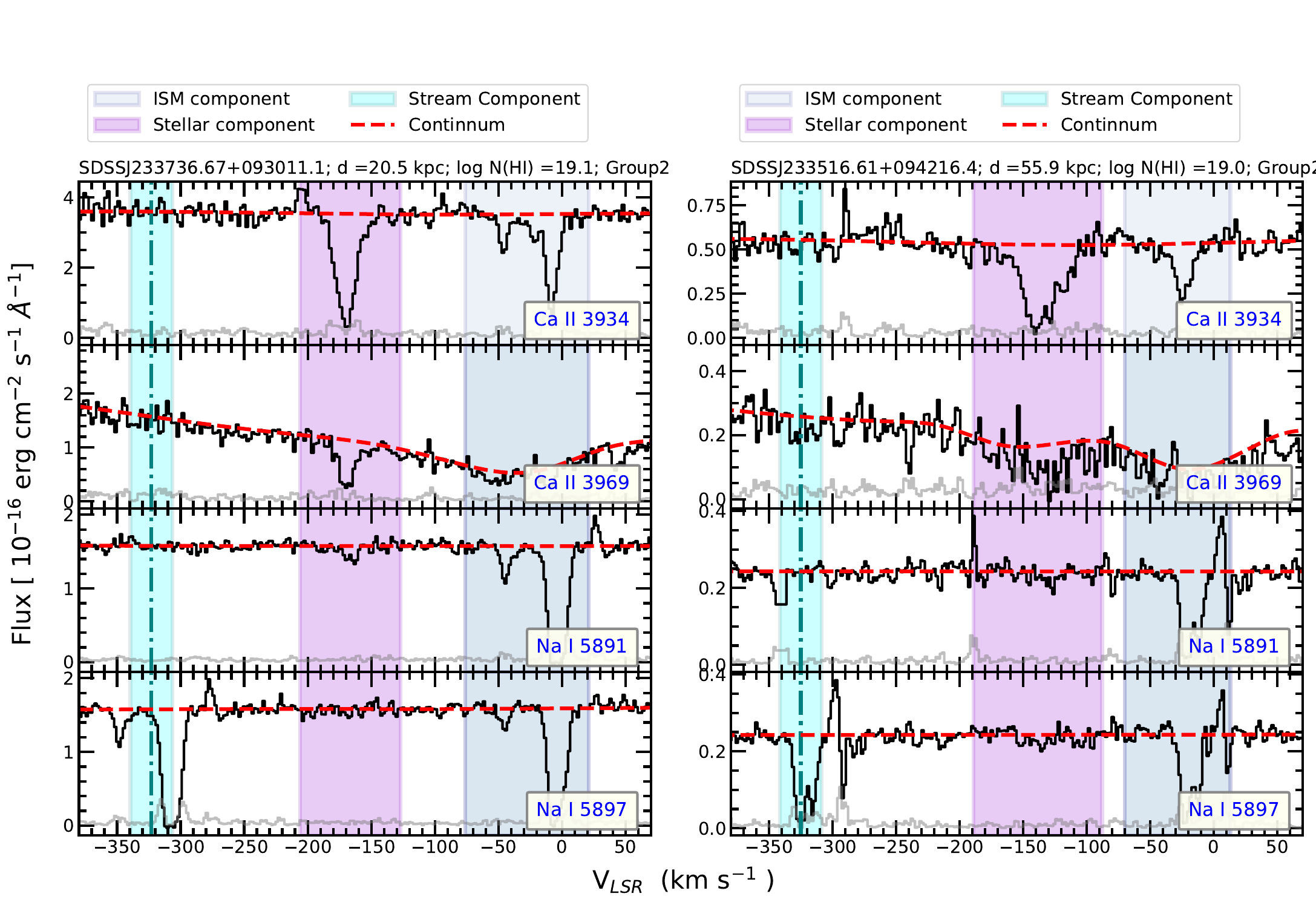}
 \caption{Same as Fig.~\ref{fig:stack_set1}, but for the Group 2 BHB stars SDSS J233736.67+093011.1 (expected Stream velocity of $\approx-$323 \kms) in the \emph{left} panel, and SDSS J233516.61+094216.4 (expected Stream velocity of $\approx-$325 \kms) in the \emph{right} panel.}
 \label{fig:stack_set3} 
\end{figure*}
%%%--------------------------

%%%%------------- App Fig. 4 stack NaI-----------
\begin{figure}[!hbt]
\hspace{-0.2in}
 \includegraphics[width=0.5\textwidth]{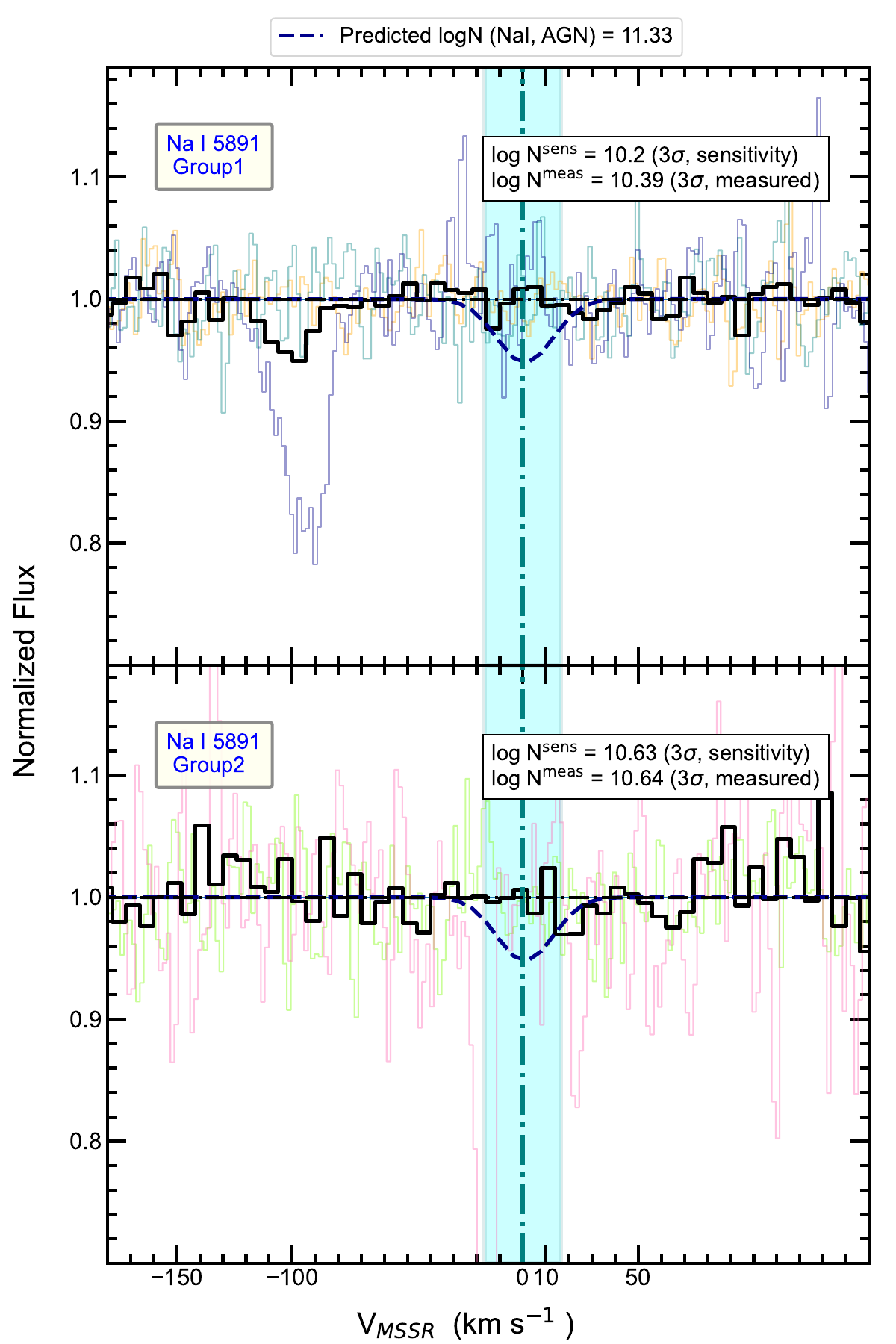}
 \caption{Same as Fig.~\ref{fig:wmean_stacks_CaII}, but for the \NaI~5891 line.}
 \label{fig:wmean_stacks_NaI} 
\end{figure}
%%%--------------

\bibliography{Reference}{}
\bibliographystyle{aasjournal}

\end{document}